\pdfoutput=1
\documentclass[twocolumn,english,aps,superscriptaddress,prb]{revtex4-1}

\usepackage{babel}
\usepackage{amsmath}
\usepackage{amssymb}
\usepackage{graphicx}

\begin{document}

\title{Exact zero modes in frustrated Haldane chains}

\author{Natalia Chepiga}
\affiliation{Institute of Physics, Ecole Polytechnique F\'ed\'erale de Lausanne (EPFL), CH-1015 Lausanne, Switzerland}
\author{Fr\'ed\'eric Mila}
\affiliation{Institute of Physics, Ecole Polytechnique F\'ed\'erale de Lausanne (EPFL), CH-1015 Lausanne, Switzerland}

\date{\today}
\begin{abstract} 
We show that the effective coupling between the spin-1/2 edge states of a spin-1 chain of finite length can be continuously tuned by frustration. 
For the $J_1-J_2$ model with nearest and next-nearest neighbor antiferromagnetic interactions, we show that the effective coupling 
in a chain of length $L$ changes sign $N\simeq 0.38 L$  times in the 
window $0.28\lesssim J_2/J_1 \lesssim 0.75$ where the short-range correlations are incommensurate. This implies that there are
$N$ 
zero modes where the singlet and the triplet are strictly degenerate, i.e. $N$ values of $J_2/J_1$ where the spin-1/2 edge states 
are completely decoupled.  We argue that this effect must be generic for all incommensurate phases with localized edge states,
and we briefly discuss a few experimental implications. 
\end{abstract}

\maketitle


Topological matter is currently attracting a lot of attention. One of the first examples is the spin-1 Heisenberg chain, which has long been known to have a finite bulk gap\cite{haldane} and spin-1/2 edge states\cite{kennedy,hagiwara}, and which has recently been shown to be an example of a symmetry-protected topological phase\cite{pollmann}. 
In one-dimensional fermionic systems with pairing, also known as the Kitaev chain\cite{kitaev}, Majorana fermions appear at the edges of a chain in the topologically non-trivial phase. The detection of the emergent Majorana fermions relies on their impact on the local tunneling density of states\cite{mourik,nadj-perge} or on the presence of two quasi-degenerate low-lying states in open systems. The presence of two
quasi-degenerate low-lying states can be most easily detected if these two states cross as a function of an external parameter, such as the chemical potential in fermionic
chains\cite{dassarma} or an external magnetic field in spin chains. Such level crossings have been recently detected in chains of Co adatoms\cite{toskovic}, and their interpretation in terms of localized
Majorana fermions worked out in details\cite{mila,vionnet}. At each level crossing, there is an exact zero mode, i.e. an excitation whose energy vanishes exactly. In the fermionic model,
the exact zero modes appear when the Majorana edge states are rigorously decoupled. It is natural to ask whether a similar effect can be induced in other topological phases
with edge states, in particular in the spin-1 chain. This would imply the presence of completely free emergent spins 1/2 at the end of a finite chain, an interesting possibility
for qubits.

In the standard spin-1 Heisenberg chain with only nearest-neighbor coupling, the spin-1/2 edge states are coupled by an effective interaction that decays exponentially with the length of the chain, and whose sign depends on the parity of the number of sites. For even chains, the coupling is antiferromagnetic, while for odd chains, it is ferromagnetic. Accordingly, the ground state is a singlet with a low-lying triplet excitation (the Kennedy triplet \cite{kennedy,hagiwara}) if the number of sites is even, while it is a triplet with a low-lying singlet if the number of sites is odd. This behavior can be traced back to the antiferromagnetic nature of the spin-spin correlations. Edge spins can be expected to be ferromagnetically aligned if the number of bonds that separate them is even, as in odd chains, while they will be antiparallel if the number of bonds is odd, as in  even chains. According to this picture, a simple way to monitor the coupling between the edge spins would be to induce incommensurate correlations. Then, for a given length, one can expect the relative orientation of the edge spins to change from parallel to antiparallel as a function of the wave vector of the fluctuations. To the best of our knowledge, in spite of decades of work on edge states in spin chains\cite{white1993,miyashita,ng1,sorensen_affleck,ng2,polizzi,ng3,eggert}, this simple possibility has not been explored. 

To implement and test this idea, let us introduce next-nearest neighbor interactions into the spin-1 Heisenberg chain. The model is defined by the following Hamiltonian:
\begin{equation}
\label{Eq:J1J2}
H=J_1\sum_{i=1}^{N-1}{\bf S}_{i}\cdot{\bf S}_{i+1}+J_2\sum_{i=2}^{N-1}{\bf S}_{i-1}\cdot{\bf S}_{i+1},
\end{equation}
where both couplings are assumed to be antiferromagnetic. Without loss of generality we set $J_1=1$. This model has been extensively studied with the density matrix renormalization group (DMRG)\cite{dmrg1,dmrg2,dmrg3,dmrg4}, and its properties are well understood\cite{kolezhuk_prl,kolezhukPRB,kolezhuk_connectivity}. For small $J_2$, the system is in the Haldane phase. At $J_2\simeq 0.75$, it undergoes  a phase transition into another gapless and translationally invariant phase known as the next-nearest-neighbor (NNN) Haldane phase in which valence-bond singlets are formed on $J_2$ bonds. This phase is topologically trivial, with no edge states, and the transition is an example of a first-order topological transition. On top of this phase transition, and most interestingly for our present purpose, there is also a disorder point in the Haldane phase at $J_2\simeq 0.28$ beyond which the short-range correlations become incommensurate and remain so up to the first-order transition. So, in view of the discussion above, we will calculate the two low-lying in-gap states of that model as a function of $J_2$, with
emphasis on the parameter range $0.28\lesssim J_2/J_1 \lesssim 0.75$.

In order to be able to study large enough systems, we have performed DMRG calculations on finite-size spin-1 chains. The determination of the two low-lying states is very easy if the singlet is the ground state since the first excitation, which is a triplet, is the ground state of the $S^z_\text{tot}=1$ sector. When the ground state is the triplet, the first excitation, which is the singlet, is the first excited state of the $S^z_\text{tot}=0$ sector. It can be accessed by a modification of the ground state algorithm to target excited states\cite{dmrg4,verstraete,dmrg5,chandross,bursill,ortolani,mcculloch}, or, as we showed recently, by keeping track of the spectrum of the effective Hamiltonian during sweeps of the ground state algorithm\cite{chepiga_dmrg}. This is the method that we have used throughout, performing up to 10 DMRG sweeps and keeping up to $D=1000$ states. 

\begin{figure}[h!]
\includegraphics[width=0.47\textwidth]{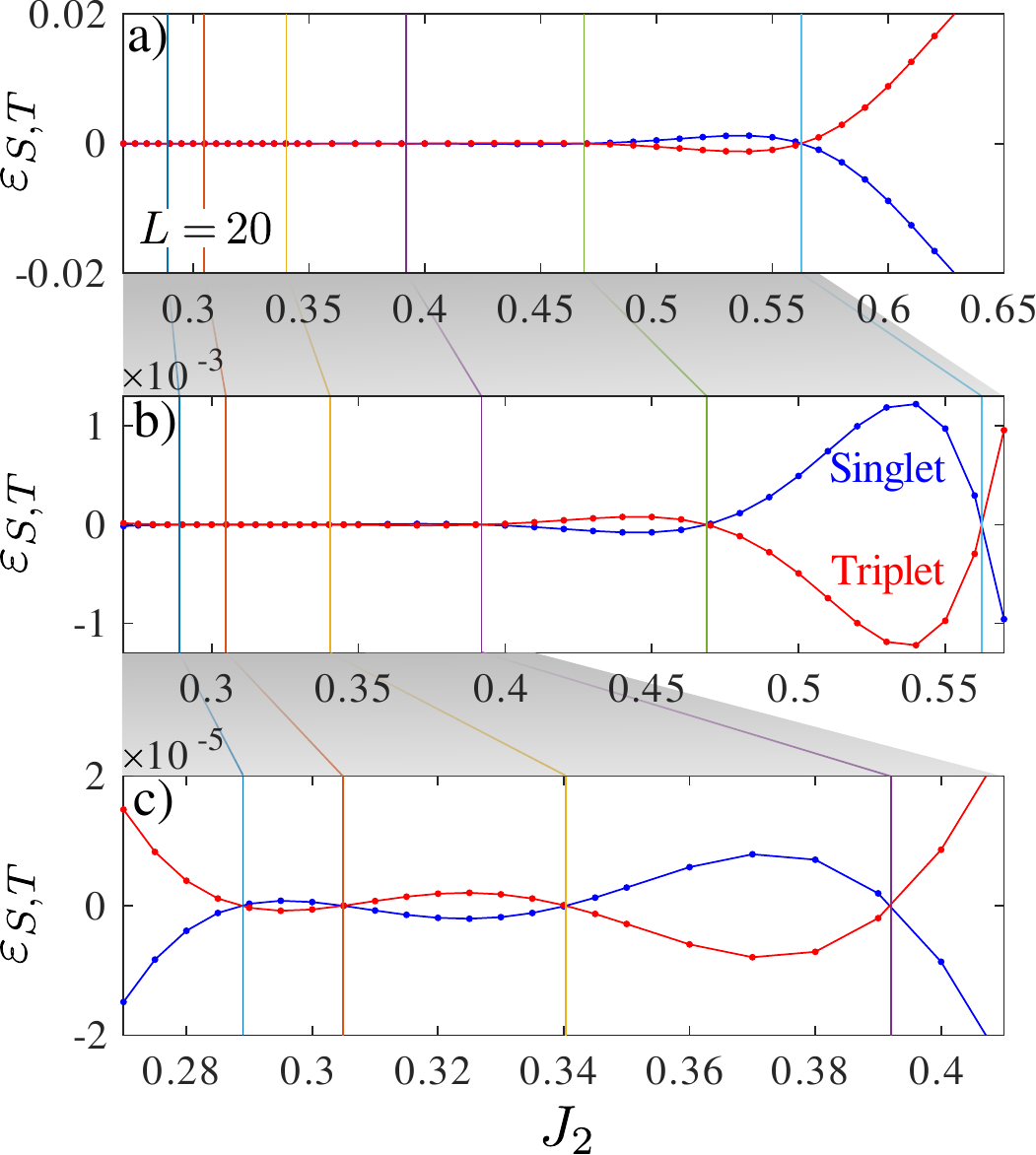}
\caption{(Color online) Multiple crossings between singlet and triplet low-lying energy levels for L=20 as a function of the next-nearest-neighbor coupling constant $J_2$. (b) and (c) are enlarged parts of (a). }
\label{fig:gap20}
\end{figure}

\begin{figure}[h!]
\includegraphics[width=0.49\textwidth]{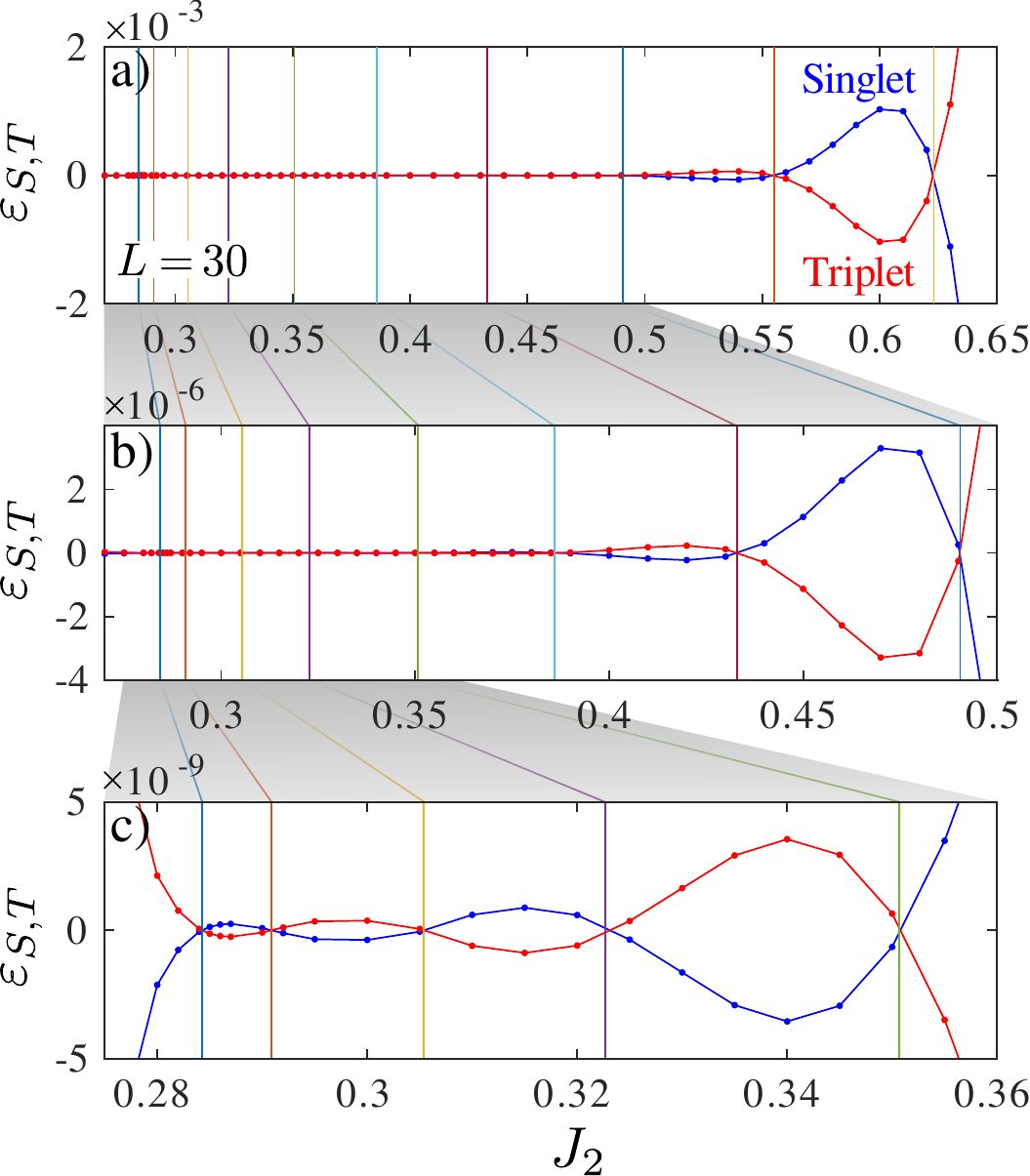}
\caption{(Color online) Same as Fig.\ref{fig:gap20}, but for $N=30$}
\label{fig:gap30}
\end{figure}

The results for 20 and 30 sites are shown in Fig.\ref{fig:gap20} and Fig.\ref{fig:gap30} respectively. Since what we are interested in is the relative position of the low-lying singlet energy $E_S$ and triplet energy $E_T$ relative to each other, we have used their average as the reference energy and plotted $\varepsilon_{S,T}=E_{S,T}-1/2(E_S+E_T)$. 
As expected, there are several level crossings in the interval where the correlations are incommensurate.  
The amplitude of the gap is extremely small close to the disorder point and increases significantly far away from it. So, in order to show all crossings, we had to repeat
the plots with different scales.
Remarkably, the number of crossings increases with the system size: There are 6 level crossings for 20 sites and 10 for 30 sites. 

\begin{figure}[h!]
\includegraphics[width=0.45\textwidth]{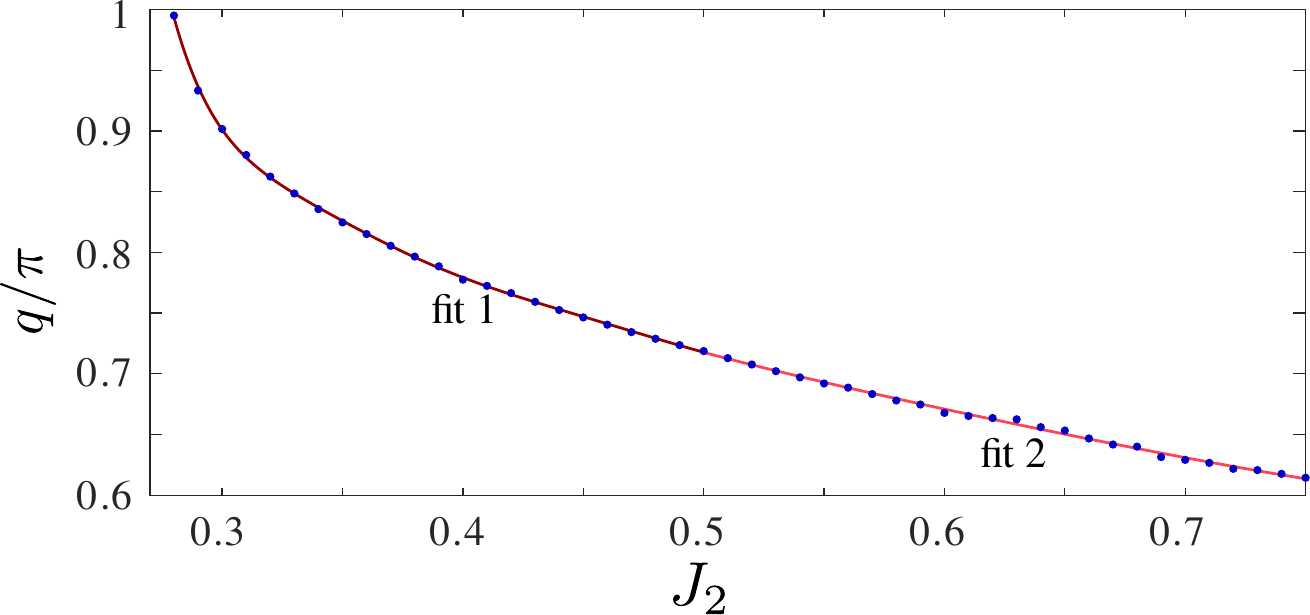}
\caption{(Color online) Wave-vector $q$ as a function of the next-nearest-neighbor interaction $J_2$. The disorder point is located around $J_2\simeq 0.28$, beyond which the Haldane phase is incommensurate.}
\label{fig:wavevector}
\end{figure}

\begin{figure}[h!]
\includegraphics[width=0.49\textwidth]{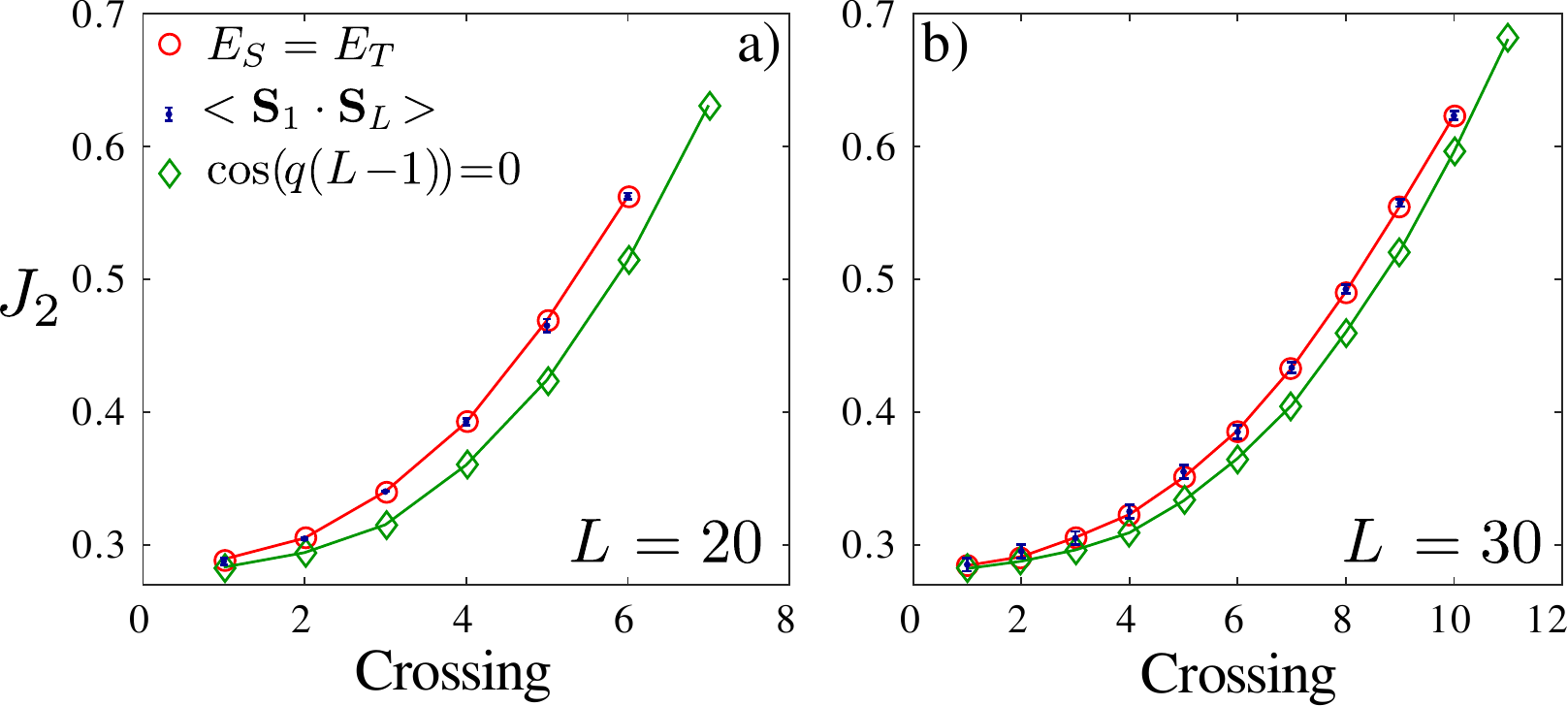}
\caption{(Color online) Location of the exact zero modes in an open chains with a) $L=20$ and b) $L=30$. Red circles stand for level crossings between the low-lying singlet and triplet energy levels, blue dots with error bars for the jumps in the spin-spin correlation between the first and last sites, and green diamond for the solutions of $\cos(q(L-1))=0$.}
\label{fig:position_of_crossings}
\end{figure}

As a first step toward the interpretation of these results, we have calculated the correlation between the first and the last spins $\langle {\bf S}_{1}\cdot {\bf S}_{L}\rangle$ as a function of $J_2$. Interestingly enough, within the error bars, these correlations change sign at the same values of $J_2$ at which the level crossings occur (see Fig.\ref{fig:position_of_crossings}).
Now, if we keep track of the wave-vector of the short-range correlations as a function of $J_2$, we should be able to predict the sign of the correlation between the first and last spin.
To extract the wave-vector
of the short-range correlations, we have used DMRG to calculate the ground-state spin-spin correlations defined by:
\begin{equation}
C_{L/2}(x)=\langle {\bf S}_{L/2}\cdot {\bf S}_{L/2+x}\rangle.
\end{equation}
It can be well fitted with the Ornstein-Zernike form\cite{ornstein_zernike}:
\begin{equation}
C_{OZ}(x)\propto \cos(q\cdot x)\frac{e^{-x/\xi}}{\sqrt{x}}.
\label{eq:OZ}
\end{equation}
The result for the wave-vector $q$ as a function of $J_2$ is presented in Fig.\ref{fig:wavevector}. In agreement with previous results for the disorder point, the wave-vector starts decreasing from $\pi$ at  $J_2\simeq 0.28$ to reach the value $q_\text{min}\simeq 0.62 \pi$ at the first-order transition boundary $J_2\simeq 0.75$. 
For further use, the discrete set of data points has been fitted by two continuous functions obtained by 8-th degree polynomial fits of the numerical data for $0.28<J_2\leq 0.5$ and $0.5\leq J_2\leq 0.75$.

Neglecting for a moment edge effects, i.e. assuming that the correlations are rigorously described by a single wave-vector $q$, the correlation between the first and last spin is proportional
to $\cos q (L-1)$, implying that it will vanish for $q=\pi/2(L-1)+k \pi/(L-1)$, $k$ integer. The corresponding values of $J_2$ are compared in Fig.\ref{fig:position_of_crossings} with those
at which the level crossings between the singlet and the triplet occur for 20 and 30 sites. The agreement is good, suggesting that this is the right physical picture.  It is not perfect however. The critical values of $J_2$ do not coincide exactly (although the agreement is already much better for 30 sites than for 20 sites), and the criterion $\cos q (L-1)=0$ gives rise to one more critical value. We can think of two explanations for theses small discrepancies. First, the edge spins are not located precisely at the edge spins, but are delocalized over the correlation length. The correlation length is small in the vicinity of the disorder point, but it increases fast close to the first order transition at $J_2\approx0.75$. So it might be better to consider an effective distance between the edge spins that decreases slightly with incresing $J_2$. Qualitatively, this goes in the right direction, but the effect is difficult to quantify.

\begin{figure}[h!]
\includegraphics[width=0.49\textwidth]{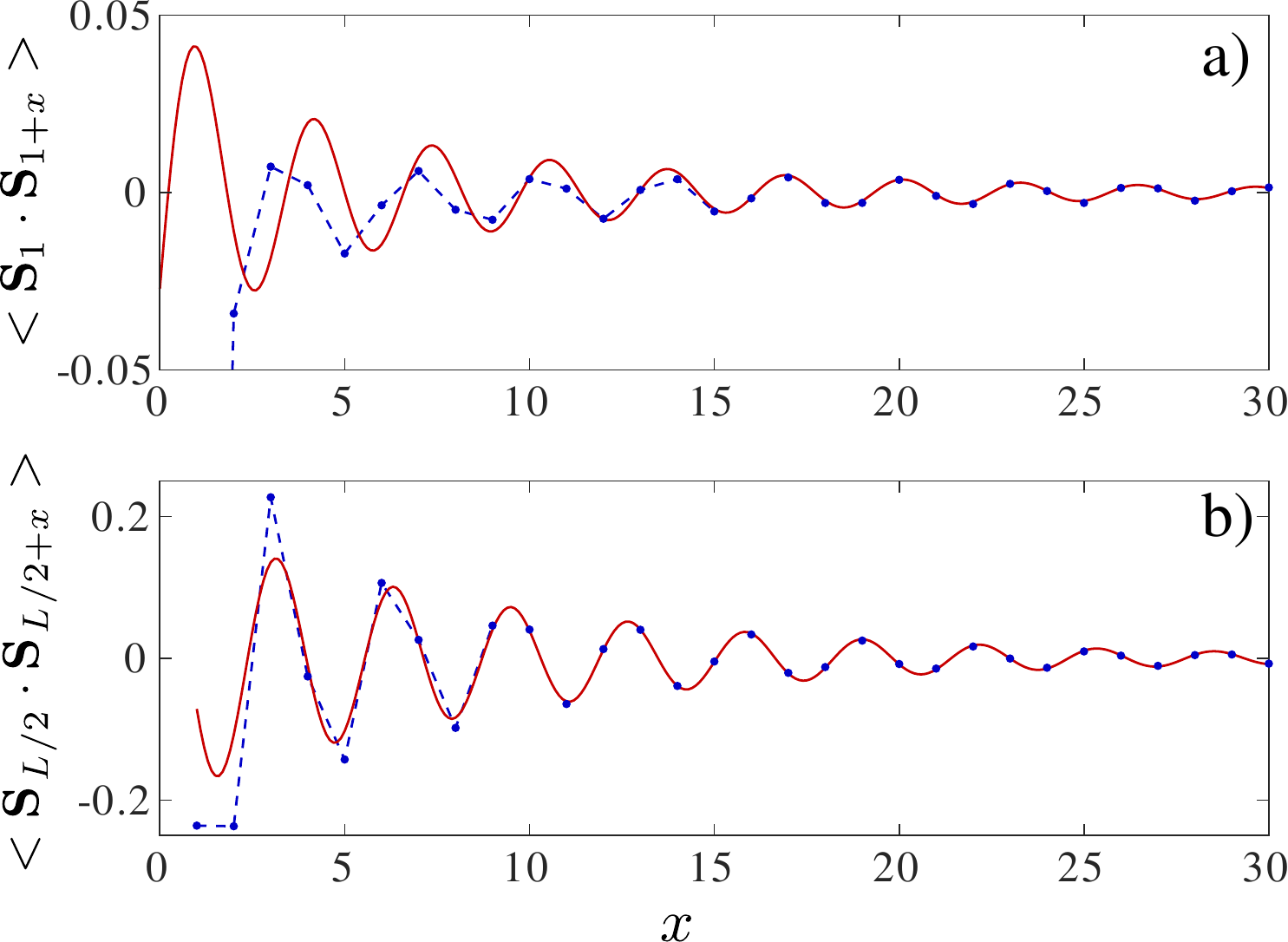}
\caption{(Color online) Spin-spin correlations starting from an edge (a) and from the middle (b) of a chain of L=150 sites deep in the incommensurate phase ($J_2=0.7$).
The red line is the result of fitting the points from 10 to 30 from the reference sites with Ornstein-Zernike formula. The discrepancy in periodicity that appears close to the
edge (a), but not close to the middle (b), shows that the wave vector decreases close to the edge.}
\label{fig:q_on_edge}
\end{figure}

The second source of discrepancy between the two curves comes from the fact that the wave number $q$ is not uniform along the chain, but decreases close to the edges, resulting again
in a shift of the critical values of $J_2$ where the correlation between the edge spins change sign. This can be deduced from Fig.\ref{fig:q_on_edge} that shows the spin-spin correlation for a) $C_{1}=\langle {\bf S}_1\cdot {\bf S}_{1+x}\rangle$ and b) $C_{L/2}=\langle {\bf S}_{L/2}\cdot {\bf S}_{L/2+x}\rangle$. In both cases we have fitted the spin-spin correlation for $10\leq x\leq 60$ with the Ornstein-Zernike form of Eq.\ref{eq:OZ}. One can notice that for small values of $x$ in Fig.\ref{fig:q_on_edge}(b) the period of the oscillations remains the same, although there is small difference in the amplitude. By contrast, in Fig.\ref{fig:q_on_edge}(a) the period of the oscillations increases for small $x$, implying that the wave vector decreases close to the edges.

\begin{figure}[h!]
\includegraphics[width=0.49\textwidth]{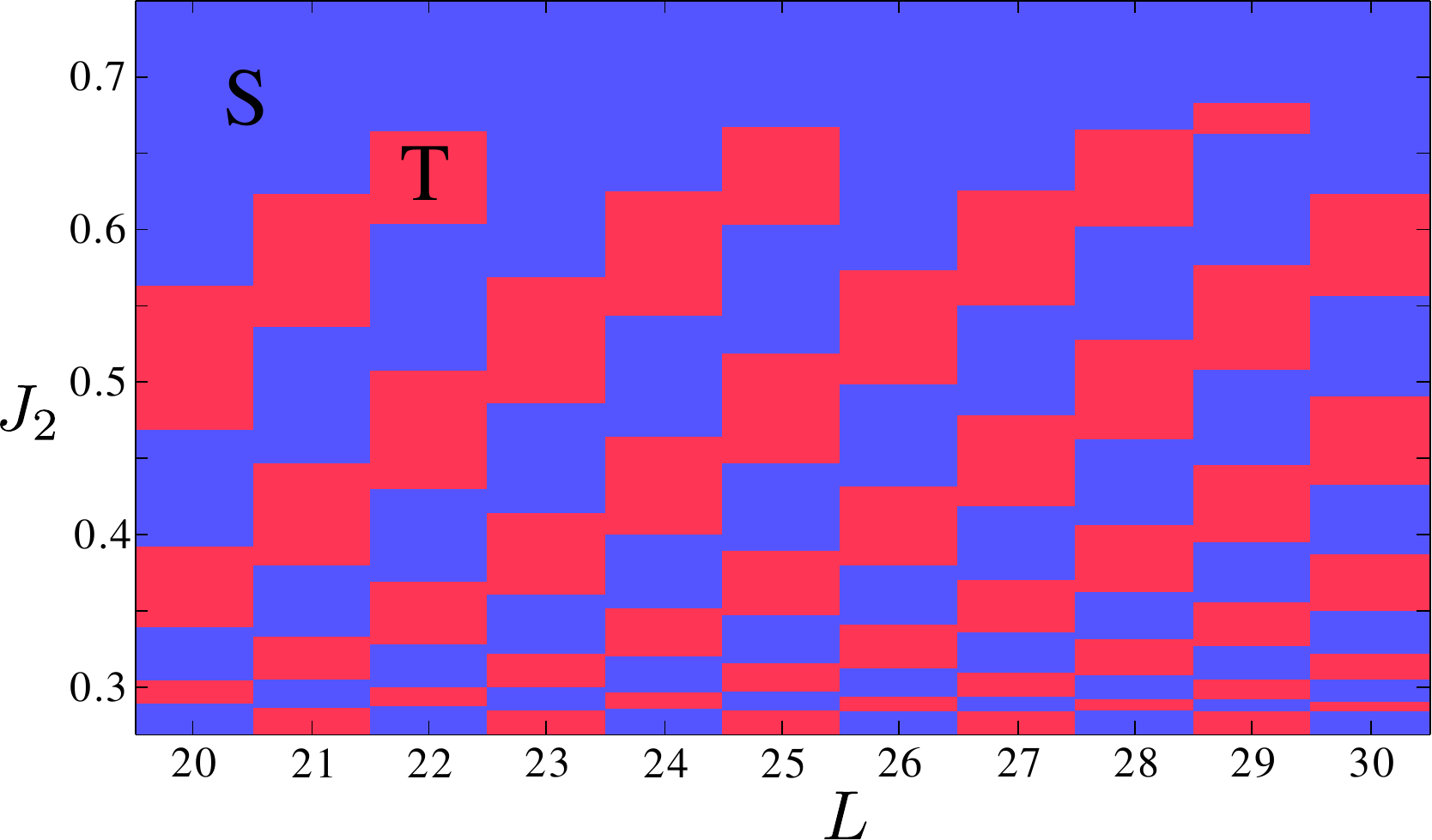}
\caption{(Color online) Phase diagram of an open chain as a function of $J_2$  and system size in the range of $20\leq L\leq 30$. Blue (red) areas stand for singlet (triplet) ground states with a triplet (singlet) low-lying state. The lower limit $J_2=0.27$ lies in the commensurate Haldane phase, where the ground state simply alternates between singlet and triplet for even and odd chains. The upper limit $J_2=0.75$ lies in the NNN-Haldane phase, where  the ground state is always a singlet.}
\label{fig:size_dependence}
\end{figure}

Still, in view of the semiquantitative agreement between the two results, we use the criterion $\cos q (L-1)=0$ to estimate the number of crossings. Since $0.62 \pi \lesssim q \leq \pi$,
the parametrization  $q=\pi/2(L-1)+k \pi/(L-1)$, $k$ integer, leads to the condition $0.62(L-1)\leq k+0.5\leq L-1$, implying that the number of crossings is approximately given by the 
integer part of $0.38(L-1)$. In particular, this simple theory predicts that the number of level crossings grows linearly with the number of sites.
This prediction agrees with the phase diagram obtained numerically for various system sizes in the range $20\leq L\leq 30$ and presented in Fig.\ref{fig:size_dependence}.  As expected, in the commensurate region of the Haldane phase ($J_2<0.28$), the ground state of an open chain is a singlet if the number of sites $L$ is even and a triplet if the number of sites is odd. Upon increasing the next-nearest-neighbor interaction, the system undergoes multiple crossings in such a way that the ground state is always in the singlet sector in the vicinity of  the phase transition between the Haldane and the NNN-Haldane phases. This constraint implies that sometimes a system with $L$ sites has one crossing less than that of size $L-1$, but 
the number of crossings is in any case equal to the integer part of $0.38(L -1)\pm1$.

Let us now discuss the implications of these results beyond the model of Eq.\ref{Eq:J1J2}. Since the main mechanism is the presence of incommensurate short-range correlations, we expect the effect to be present in the entire incommensurate phase that has been found in two generalizations of the model of Eq. \ref{Eq:J1J2}, the $J_1-J_2-J_\text{biq}$ model that contains an additional biquadratic interaction\cite{pixley,chepiga_biq}, and the  $J_1-J_2-J_3$ model that contains an additional three-site interaction\cite{chepigaRC,chepigaPRB}. More generally, we expect the effect to be present
whenever incommensurate correlations develop in a one-dimensional topological phase with edge states. In fact, the level crossings that have been recently discussed in the transverse
field Ising model with an additional longitudinal coupling, or equivalently in the Kitaev chain when the pairing amplitude is smaller than the hopping term, have been explained in terms
of oscillations of the Majorana edge state wave functions\cite{vionnet}, in a parameter range where incommensurate spin-spin correlations are indeed present\cite{mccoy}.
Let us note to be complete that these zero modes do not appear to be strong zero modes  in the terminology of Paul Fendley\cite{fendley}. The spectrum does not seem to consist of quasi-degenerate pairs of state, unlike that of e.g. the XYZ chain, but only of one pair of low-energy quasi-degenerate singlet and triplet states.

This new instance of exact zero modes might have interesting experimental implications. Indeed, if a system
lies in the incommensurate Haldane phase with e.g. a large enough $J_2$, it should be possible to go from decoupled to ferromagnetically or antiferromagnetically coupled spin-1/2 edge 
states by a tiny modification of the geometry of the system since only a very small change of the $J_2/J_1$ ratio around one of the critical values will be required. In
bulk compounds, such modifications can be achieved by applying pressure. Inspired by the current activity on the magnetism of adatoms\cite{ternes}, we can also think of more direct implementations.
A zig-zag chain of spin-1 adatoms would be a natural realization of the $J_1-J_2$ model. A simple way to demonstrate the effect described in the present paper would be to show
that the even-odd rule is violated in the presence of frustration, for instance that a system with an even number of sites has a triplet ground state. The triplet character of the 
ground state is easily accessible since edge states lead, in the triplet state, to a very specific local magnetization pattern\cite{white1993,miyashita,sorensen_affleck,polizzi} that can be probed by STM experiments\cite{fernandez}. One can also imagine to modify the local geometry by bending the substrate. This could be a way to control the coupling of spin-1/2 edge states of finite segments seen as qubits. Finally, edge states are also accessible in ring-shaped molecular magnets\cite{affronte}. Developing further these ideas goes beyond the scope of the present paper however and is left for future investigation.

We thank Ian Affleck and Paul Fendley for useful discussions on related topics. This work has been supported by the Swiss National Science Foundation.

\bibliographystyle{apsrev4-1}
\bibliography{bibliography}

\end{document}